# *Ex vivo* estimation of cementless femoral stem stability using an instrumented hammer

Hugues Albini Lomami[a,b], Camille Damour[a], Giuseppe Rosi[a], Anne-Sophie Poudrel[a], Arnaud Dubory[b,c], Charles-Henri Flouzat-Lachaniette[b,c], Guillaume Haiat[a]

[a]CNRS, Laboratoire de Modélisation et de Simulation Multi-Echelle, UMR CNRS 8208,61 Avenue du Général de Gaulle, Créteil 94010, France
[b]INSERM U955, IMRB Université Paris-Est, 51 avenue du Maréchal de Lattre de Tassigny, 94000 Créteil, France.
[c]Service de Chirurgie Orthopédique et Traumatologique, Hôpital Henri Mondor AP-HP, CHU Paris 12, Université Paris-Est, 51 avenue du Maréchal de Lattre de Tassigny, 94000 Créteil, France.

**Abstract-**

*Background*

The success of cementless hip arthroplasty depends on the primary stability of the femoral stem. It remains difficult to assess the optimal number of impacts to guarantee the femoral stem stability while avoiding bone fracture. The aim of this study is to validate a method using a hammer instrumented with a force sensor to monitor the insertion of femoral stem in bovine femoral samples.

*Methods*

Different cementless femoral stem were impacted into five bovine femur samples, leading to 99 configurations. Three methods were used to quantify the insertion endpoint: the impact hammer, video motion tracking and the surgeon proprioception. For each configuration, the number of impacts performed by the surgeon until he felt a correct insertion was noted $N_{surg}$. The insertion depth $E$ was measured through video motion tracking, and the impact number $N_{vid}$ corresponding to the end of the insertion was estimated. Two indicators, noted $I$ and $D$, were determined from the analysis of the time variation of the force, and the impact number $N_d$ corresponding to a threshold reached in $D$ variation was estimated.

*Findings*

The pullout force of the femoral stem was significantly correlated with $I$ ($R^2$=0.81). The values of $N_{surg}$, $N_{vid}$ and $N_d$ were similar for all configurations.

*Interpretation*

The results validate the use of the impact hammer to assess the primary stability of the femoral stem and the moment when the surgeon should stop the impaction procedure for an optimal insertion, which could lead to the development of a decision support system.

**Keywords:** Hip arthroplasty ; Cementless femoral stem ; Primary stability ; Impact hammer

# 1. Introduction

Cementless hip arthroplasty is now commonly performed in clinical practice to replace defective hip joints. Prior to the femoral stem (FS) insertion into the femoral shaft, the femur is reamed with a rasp that is slightly smaller than the size of the FS following a press-fit procedure, in order to obtain an adapted primary implant stability. The FS primary stability into the host bone is an important determinant of the surgical success and the biomechanical properties of the bone-implant interface (BII) plays a crucial role in osseointegration phenomena that determines secondary stability. Micromotion at the BII should be minimized in order to avoid fibrous tissue formation that may lead to aseptic loosening. However, excessive stresses at the BII should be avoided since it may lead to bone necrosis and/or to femoral fracture.

In order to obtain an optimal stability of the FS and to avoid the aforementioned problems, the surgeon must adapt the number and the energy of the impacts realized to insert the FS into the femoral shaft, which is done using an orthopedic hammer. On the one hand, the FS should be completely inserted into the femoral shaft in order to maximize the bone-implant contact. On the other hand, impacts with too much energy may lead to intraoperative and/or post-operative peri-prosthetic femoral fracture. Surgeons use their proprioception to determine when impacts on the FS should be stopped, by listening to the sound of the hammer impact on the ancillary (Sakai et al., 2011; Morohashi et al., 2017). However, such criterion remains qualitative and highly depends on the surgeon's perception and experience.

A few quantitative methods aiming at assessing the insertion of the FS into the host bone have been reported in the literature. One of them consists in analyzing micromotions at the BII using linear variable differential transducers (LVDTs). This method was used to compare the FS stability in bone mimicking phantoms using different kinds of stem fixation (Baleani et al., 2001), for resurfacing hip implants with different loading conditions (Cristofolini et al., 2006), to study the effects of implant undersizing (Fottner et al., 2017), and by considering unidirectional micromotion (Leuridan et al., 2017). Other studies investigated the behavior of FS with different sizes or shapes in human femur (Bieger et al., 2013) and in composite bone (Schmidutz et al., 2017). Nonetheless, due to complex experimental settings, it remains difficult to employ LVDTs in clinical practice.

Another technique consists in using vibrational analysis to follow the FS insertion (Denis et al., 2017). This method has been validated in artificial bones (Pastrav et al., 2008)(Pastrav et al., 2009) and in cadaveric specimens (Varini et al., 2010). Although such approach is promising, much work remains to be done for clinical implementation.

Micro-computed tomography (µCT) has been used to quantify the micromotions at the BII (Gortchacow et al., 2012). This technique was studied to assess FS primary stability in cadaveric femurs (Malfroy Camine et al., 2018). However, metallic implants cause artefacts, thus limiting the quantitative investigation of the BII (Park et al., 2005). The use of synchrotron radiation µCT has been suggested (Müller, B et al., 2004; Neldam and Pinholt, 2014), but this technique cannot be employed in clinical practice.

A method based on the analysis of the variation of the force applied between an instrumented hammer and the ancillary during impacts was developed by our group primarily in the context of the assessment of the acetabular cup (AC) implant stability. This approach was first assessed using mass drops, which validated the possibility of monitoring AC cups insertion in bovine bone samples by analyzing the behavior of the contact duration (Mathieu et al., 2014, 2013). Another indicator based on the impact momentum was used to assess the AC implant primary stability (Michel et al., 2015). The method was validated *ex vivo* (Michel et al., 2016a) and in anatomical subjects (Michel et al., 2016b). An alternative hammer instrumented by strain sensors was also developed to assess the AC implant stability (Tijou et al., 2017). In order to better understand the phenomena related to the AC insertion, static (Nguyen et al., 2017) and dynamic (in the time (Michel et al., 2017) and frequency (Henyš et al., 2018) domains) finite element models (FEM) were developed. Another group has used a comparable approach using a force cell instrumented hammer to measure an impaction energy on AC implants (Henyš and Čapek, 2018). More recently, this same technique using an instrumented hammer was applied to follow the FS insertion in bone mimicking phantoms (Tijou et al., 2018). Another indicator $D$ based on the time history of the force recorded by the hammer during successive impacts was used to estimate the FS insertion endpoint. However, the *in vitro* study implied using bone mimicking phantoms, which have different material properties compared to bone tissue.

The aim of the present study is to validate a method based on the analysis of the variation of the force as a function of time obtained during impacts realized between an instrumented hammer and a FS inserted in bovine femoral samples. The signal processing technique aims at quantifying the insertion endpoint of the FS using three methods: i) the instrumented hammer, ii) video motion tracking and iii) the surgeon proprioception. To do so, the instrumented hammer was used to insert different sizes of FS in five bovine femurs. A system using optical markers was used to follow the FS insertion with video motion tracking (VMT). The orthopedic surgeon determined the FS insertion endpoint using his proprioception. Eventually, the impact momentum was determined and compared with the FS pullout force.

# 2. Materials and methods

## 2.1. Implants and specimens

Cementless FS (CERAFIT R-MIS) manufactured by Ceraver (Roissy, France) were used. These implants were made of titanium alloy (TiAl6Al4V) and coated by hydroxyapatite to enhance osseointegration phenomena. Seven sizes were tested, from size 7 to 13, with their corresponding rasps also manufactured by Ceraver (Roissy, France) used prior to FS insertion. An ancillary was screwed into the FS so as to obtain a rigid bilateral fixation between the two parts.

Five bovine femurs were collected from the local butcher's shop and were prepared similarly to (Tijou et al., 2018). The sample was cleaned from all residual soft tissues. Each sample was cut at the boundary between the diaphysis and the distal epiphysis, lower than the middle of the femur, so that the biggest implant could be fully inserted in trabecular bone. The distal end of the sample was then embedded in a fast-hardening resin (SmoothCast 300 polymer, Smooth-On, Easton, USA). Eventually, an osteotomy was performed on the femoral neck according to the clinical procedure. Figure 1 shows an image of the sample.

## 2.2. Experimental set-up for impacts

### 2.2.1. Femoral stem insertion

A schematic representation of the experimental set-up is shown in Fig. 1. The resin block was clamped inside a vise in order to avoid any displacement perpendicular to the femur axis. The 1.3 kg hammer used to insert the FS inside the sample was the same as in our previous study (Tijou et al., 2018). A dynamic piezoelectric force sensor (208C05, PCB Piezoelectronics, Depew, New York, USA) with a measurement range up to 22 kN in compression was screwed in the center of the hammer impacting face.

s(t) was defined as the variation of the force (given in kN) applied to the ancillary and recorded by the force sensor as a function of time $t$ (given in ms). A data acquisition module (NI 9234, National Instruments, Austin, TX, USA) with a sampling frequency of 51.2 kHz and a resolution of 24 bits was linked to a LabView interface (National Instruments, Austin, TX, USA) to record *s(t)* for a duration of 13 ms.

### 2.2.2. Video motion tracking

Video tracking was carried out to follow the insertion of the FS into the femoral shaft during the impaction procedure. A camera (Powershot SX410 IS, Canon, Tokyo, Japan) was used, with a frame rate of 24 images per second and a resolution of 1280x720 pixels. Optical markers were glued on the FS, the ancillary and the femur, as shown in Fig. 1. Their relative displacement was calculated with the software Tracker (Cabrillo College, Aptos, CA, USA). The following distances were defined by analyzing the image after a given impact #i by using the markers on the FS and the ancillary (which is constant throughout the experiment) as a scale to convert pixels into centimeters. $d(i)$ (given in centimeter) was defined by the distance between the markers located on the bone sample and on the ancillary after impact #*i*. Therefore, $d(0)$ corresponds to the initial distance between these same markers (before impaction). The indicator $E$ (given in centimeter) was defined by the relative motion of the FS compared to the femoral bone compared to the initial position following:

$$E(i) = d(0) - d(i) \text{ (in cm)} \tag{1}$$

### 2.2.3. Femoral stem stability assessment

After each test, the FS was pulled out from the bone sample in the direction of the axis of the stem, using a tensile testing machine (DY25, Adamel Lhomargy, Roissy en Brie, France). The resin block was fixed and a constant displacement rate of 5 mm/min was applied to a hook screwed to the FS at the ancillary location and attached to the machine. The pullout force *F*, given in daN, defined as the maximum value of the measured force, was determined for each test and corresponds to the FS primary stability.

## 2.2.4. Signal processing

A dedicated signal processing method was implemented in order to analyze the signals *s(t)*, as illustrated in Fig. 2. Two different indicators were calculated for each impact.

The first indicator, noted *D* and given in ms, was defined as the time difference between the time of the second and the first local maxima of the signal *s(t)*, and was calculated as follows:

$$D = f_2(s) - f_1(s) \text{ (in ms)} \tag{2}$$

where the functions $f_1$ and $f_2$ are applied to a signal *s(t)* and were defined by the time of the maximum value of the first and second peak of the signal s(t), respectively.

The second indicator, noted *I* and given in kN, corresponds to the impact momentum, and was defined similarly to our previous works (Tijou et al., 2018) as:

$$I = \frac{1}{t_2 - t_1} \int_{f_1(s)+t_1}^{f_1(s)+t_2} s(t)\, dt, \text{ (in kN)} \tag{3}$$

where $t_1$= 0.08 ms and $t_2$= 0.35 ms delimit the integration bounds. The choice of the time window will be discussed in section 4.

## 2.2.5. Experimental protocol

Two series of experiments were performed by a trained orthopedic surgeon. In the first series (similar to what was done previously in (Tijou et al., 2018), see Fig. 3), all FS were inserted in a cavity that was reamed with the size of the rasp corresponding to the FS, while in the second series (see Fig. 4), which aimed at obtaining relatively unstable configurations, the size of the FS was smaller than the rasp size used to ream the cavity. Four bovine samples were considered for the first series, while only one bovine sample was used for the second series.

In the first series of experiments, a cavity was initially created in each bovine bone with the smallest available rasp (size n=7). Then the impaction procedure was performed following three steps. First, the FS - ancillary system was impacted with the instrumented hammer until the surgeon considered that the FS was stable. The corresponding number of hits was noted $N_{surg}$. Secondly, 12 additional impacts with a first-peak amplitude comprised between 1 and 12 kN were carried out in order to assess whether the implant could be further inserted into the femoral shaft, and to determine the average value of the indicator *I* obtained for stable conditions of the FS. The choice of the limit values of the first peak amplitude (1 - 12 kN) and the number of additional impacts (12) will be discussed in section 4. For each impact, the indicators *D* and *E* were determined as described above. Eventually, the implant was pulled out axially and the pullout force *F* was measured.

The aforementioned procedure was repeated four times with the same implant as far as i) the sample was not fractured and ii) the surgeons felt that it was possible to obtain a good FS stability. Then, the surgeon checked if the femur was fractured. If so, the tests were stopped for the sample. If not, a bigger size of rasp and the corresponding FS was used.

The protocol followed for the second series is described in Fig. 4. In order to obtain looser fixations, the size *m* of the femoral stem must always be inferior to the size *n* of the reamer. We started the procedure with a reamer of size *n=11* and a femoral stem of size *m=n-1*. For the same cavity created by a reamer size *n*, we carried out the femoral stem insertion 4 times. Then, we checked the level of instability by measuring the pullout force.

- First, if the pullout force obtained is superior to 65 N after the fourth test, the configuration was not considered as unstable with the femoral stem size m=n-1 and we updated the size m of the rasp to m=m-1 until m=n-3.

- Second, if the pullout force obtained is inferior to 65N, the configuration was considered unstable and another configuration was carried on with a n=n+1 reamer size starting with a m=n-1 femoral stem size.

The procedure was stopped either when we reached n = 13 (because we did not have a reamer size greater than 13) or when the sample was fractured. The three-step impaction procedure was identical to the one applied in the first series of experiments. Each procedure was repeated four times.

## 2.2.6. Post-processing and data analysis

Following the experimental protocol described in the last subsection, three different analyses were performed for each insertion procedure in order to estimate the required number of impacts to reach the insertion endpoint, corresponding to the end of the migration phase of the FS into the femur.

The first method consists in following the variation of the indicator *D* as a function of the impact number *i* noted *D(i)*. For each impaction procedure, the parameter $N_d$ was defined as the number of the first impact for which:

$$D(i) \leq D_{th}, \tag{4}$$

where $D_{th}$ is a threshold chosen equal to 0.47 ms. The choice of this parameter will be discussed in section 4.

The second method uses the variation of *E* as a function of the impact number *i*, noted E(i). For each impaction procedure, the parameter $N_{vid}$ was defined as the lowest impact number that respects:

$$E(N_{vid}) \geq E_m - \delta \times E_{sd} \tag{5}$$

where $E_m$ and $E_{sd}$ are respectively the average and the standard deviation of the values of E for the last ten impacts and $\delta$ is a parameter empirically chosen equal to 7. Therefore, the parameter $N_{vid}$ corresponds to the impact number where the value of E becomes relatively close to its endpoint value given by $E_m$. The choice of $\delta$ value will be discussed in section 4.

The third method corresponds to the one used in the clinic and is based on the surgeon proprioception, as described in section 2.2.5. When the surgeon felt that the insertion endpoint of the FS was reached, the parameter $N_{surg}$ was noted.

To summarize, $N_d$, $N_{vid}$, $N_{surg}$ correspond to three values of insertion endpoint obtained for each insertion procedure using the instrumented hammer, the video motion tracking and the surgeon proprioception, respectively. In order to determine the accuracy of the three aforementioned approaches, it was necessary to quantify the difference obtained between the different parameters values $N_d$, $N_{vid}$, and $N_{surg}$. To do so:

- $M_d = N_d - N_{surg}$ was defined as the difference between the insertion endpoint obtained with the instrumented hammer and the surgeon proprioception,
- $M_{vid} = N_{vid} - N_{surg}$ was defined as the difference between the insertion endpoint obtained with the video motion tracking and the surgeon proprioception,
- $M_c = N_{vid} - N_d$ was defined as the difference between the insertion endpoint obtained with the video motion tracking and the instrumented hammer.

Subsequently, for each impaction procedure, the average and standard deviation values of the indicator *I* (noted $I_m$ and $I_{sd}$, respectively) were calculated. To obtain comparable results, only the six signals having their first-peak amplitude closest to 4500 N were selected. For each configuration, the standard deviation was calculated as the mean value of two batches of three signals, following the procedure used in our previous study (Tijou et al., 2018). Choices of the number of signals and the central first peak amplitude will be discussed in section 4.

# 3. Results

## 3.1. Stable configurations (first series)

Table 1 shows the number of impaction procedures considered for each bovine bone sample for the first series of experiments. For sample #1 and implant size equal to 8 and for sample #3 and implant size equal to 8, only three configurations were considered because of a loss of stability obtained after three configurations, which made it difficult to exploit the fourth configuration. This loss of FS stability may be due to a modification of bone properties due to repeated impacts. For all other samples and implant sizes for both series of experiments, four configurations were considered.

Figure 5 shows different examples of force signals *s(t)* obtained for six impacts of a given configuration (Sample #4, Implant size #9, cavity #9, test #3) corresponding to the first series of experiments. The respective number of each impact is indicated above each corresponding second peak of the signal. The value of *D(i)* first decreases with *i* and then stays constant after the fifth impact. Note that the shape of the signal corresponding to the insertion endpoint (n=5) changes compared to the signals obtained earlier in the insertion protocol.

The variation of the two parameters *D* and *E* as a function of the impact number *i* obtained for the same configuration as the one shown in Fig. 5 is shown in Fig. 6. The horizontal back dashed line represents a threshold penetration value equal to $E_m - \delta \times E_{sd}$ and the horizontal grey dashed line represents the threshold $D_{th}$=0.47 ms. The vertical dashed lines represent the different parameters values, $N_d$, $N_{vid}$, and $N_{surg}$. For this configuration, $N_{surg}=N_{vid}=N_d=5$, so $M_{vid}=M_d=M_c=0$.

## 3.2. Unstable configurations (second series)

Table 2 shows the number of impaction procedures considered for each bovine bone sample for the second series of experiments. Four configurations were considered for all implant and reamer sizes.

The variation of the two parameters $D$ and $E$ as a function of the impact number $i$ obtained for the implant size equal to 7 and reamer size equal to 7 is shown in Fig. 7. The dark vertical line represent the value obtained for $N_{surg}$ and $N_d$ and the clear vertical line represent the value obtained for $N_{vid}$ which is equal to 7. For this configuration, $N_{surg}=N_d=8$ and $N_{vid}=7$, so $M_{vid}=-1$; $M_d=0$; $M_c=-1$.

## 3.3. Combining stable and unstable configurations

When pooling the first and second series of experiments, the total number of configurations was equal to 99.

The distribution of the values of $M_d$, $M_{vid}$ and $M_c$, corresponding to the difference between the different methods developed for the estimation of the FS insertion endpoint is shown in Fig. 8. The average and standard deviation values of $M_d$, $M_{vid}$, and $M_c$ obtained for the 99 configurations are shown in Table 3. They are all in the same range of variation.

The relationship between the indicator $I_m$ and the pullout force $F$ is shown in Fig. 9 together with a linear regression analysis. A significant correlation is obtained between $I_m$ and $F$ and the determination coefficient $R^2$ is equal to 0.81. The errorbars correspond to the standard deviation $I_{sd}$ obtained for each configuration.

# 4. Discussion

The main purpose of the present study was to determine whether an instrumented hammer could be used to follow the femoral stem insertion and determine its endpoint in bovine bone samples. This approach has previously been studied by our group with artificial bone samples (Tijou et al., 2018) with the same hammer and the originality of the present study is to consider bone tissue that has different material properties compared to bone mimicking phantoms.

Figures 6 and 7 indicate that the indicator $D$, corresponding to the time difference between the first and the second peak of the signals, decreases as a function of the number of impacts, until it reaches a constant value. The decrease of $D$ as a function of the number of impacts can be explained to the increase of the bone-implant contact ratio during the insertion of the FS, resulting in an increase of the stiffness of the bone-implant system that may in turn explain the increase of its resonance frequency. Note that other authors (Oberst et al.,

2018)(Henyš et al., 2018)(Henyš and Čapek, 2018) also found a similar evolution of the resonance frequency during the insertion, with a comparable approach, using force signal tracking.

The results show that for the first series of experiments corresponding to stable configurations, the values of the pull-out force is comprised between around 100 and 250 daN, which is lower but of the same order of magnitude than values obtained with cadaveric specimen (Barr et al., 2015; Grechenig et al., 2015). The lower values of pull-out force obtained herein may be explained by the fact that i) the geometrical shape of bovine femoral shaft is not adapted to the shape of the FS and ii) the fact that repeated insertion may lead to modifications of the bone properties. In the present study, we considered unstable situations (the second series of experiments) in addition to the relatively stable ones in order to obtain a wide range of FS stability.

Figure 9 shows a significant correlation between the indicator $I_m$ and the pullout force $F$, for 99 configurations. The determination coefficient equal $R^2=0.81$ is in agreement with the results obtained *in vitro* ($R^2=0.67$). The present study confirms that the previous results are still valid when considering bone tissue instead of polyurethane foam.

Figure 8Figure 8 shows a good agreement between the three methods developed to assess the number of impacts required to reach the FS insertion endpoint. Again, these results are consistent with the previous *in vitro* results (Tijou et al., 2018). However, there remain some discrepancies between the methods, as shown in 8 and Table 3. The average value of $M_c$, equal to -0.19, shows a good agreement between the results obtained with VMT ($N_{vid}$) and the instrumented hammer ($N_d$). However, the average values of $M_{vid}$ (-1.17) and $M_c$ (-0.98) indicate that the surgeon evaluation of the insertion endpoint is assessed in average one impact after the results obtained with VMT and the instrumented hammer. Several factors can explain these differences. First, the surgeon proprioception is mainly qualitative and depends on the surgeon's tactile feelings, hearing and vision. Second, errors may occur using VMT due to i) changes of angular position, ii) possible macroscopic 3D movements and plastic bone deformation, iii) errors based of the image processing, such as the precision of markers tracking, which can be observed in the fluctuation of $E$ indicator even after reaching stability (Fig.6).

In this study, several parameters were chosen empirically. First, the number of additional impacts after $N_{surg}$ was chosen equal to twelve, resulting from a compromise between i) a sufficiently high number of impacts to obtain a reproducibility of the indicator $I$ and a convergence in the variation of the indicators $D$ and $E$ and ii) a sufficiently low number to minimize fracture risk. Second, the upper limit of the range of variation of the maximum force of the impacts performed after $N_{surg}$ [1-12 kN] was chosen sufficiently low to minimize fracture risk. Note that this range of variation is similar to the usual range applied by the surgeon during the insertion, before $N_{surg}$. Third, the choice of the threshold $D_{th}$ = 0.47 ms was made after analyzing the overall values of D as a function of the first peak amplitudes for impact after $N_{surg}$+1. Figure 10 shows the variation of the values of D obtained for the impacts performed after $N_{surg}$. The dotted line corresponds to the threshold $D_{th}$ chosen equal to 0.47 ms. No significant variation of $D$ was observed as a function of the amplitude of the first peak. Moreover, the value of $D$ was higher than $D_{th}$ for only 4 impacts out of 1822, which validates the approach and explains the choice of $D_{th}$.

Fourth, the value of $\delta$ was chosen empirically to find a compromise between a sufficiently high value to account for the constant increase of $E$ after $N_{surg}$ (due to plastic deformation of the bone phantom) and a sufficiently low value in order not to underestimate the value of $E_m$. Fifth, the values of $t_1$ and $t_2$ used to determine the indicator $I$ result from an optimization study and maximize the determination coefficient of the linear regression between $I_m$ and $F$. Varying the value of $t_1$ between 0.07 ms and 0.12 ms and the value of $t_2$ between 0.30 ms and 0.35 ms did not alter significantly the results, with a difference of less than 1% in $R^2$ value. Furthermore, the number of impacts considered to determine $I_m$ (equal to six) was chosen sufficiently high to obtain a consistent value of $I_m$ and standard deviation. These selected impacts corresponded to the impacts with a first peak amplitude closest to the overall median value equal to 4.5 kN.

This study presents several limitations. First, the biomechanical properties of the pelvic bone and of bovine femoral bone are different, as shown in Table 4 that compares results obtained in different studies (Michel et al., 2016b)(Chiba et al., 2011). Although the variation range of bone volume over total volume ratio (BV/TV), trabecular spacing (Tb.Sp) and structure model index (SMI) are similar, there are significant differences for trabecular thickness (Tb.Th) and trabecular number (Tb.N). Furthermore, bovine femur has a much superior thickness than human femur and thus a different quantity of surrounding bone tissue, which can lead to discrepancies. These differences justify the need to compare our results in different bone models: artificial and bovine bone samples and in cadaveric specimen in the future. Note that we did not use human cadaveric samples because the ethical committee requires to validate the method in bovine bone samples before using cadaveric samples.

Second, the multiple use of the bovine specimen is not representative of clinical situations and could lead to changes of bone properties. In particular, the same FS was inserted consecutively four times in the same bovine sample without modifying the cavity, which constitutes a limitation because the bovine sample may deform inelastically and therefore may lead to a modification of the FS seating. However, we checked that the averaged relative variation of the pull-out force during the four impaction procedures realized with the same sample was equal to 6 daN, which is significantly lower than the value of the pull-out force. Moreover, a modification of the FS seating would not affect the reliability of our approach because it would correspond to a modification of the pull-out force that is likely to be detected by the instrumented hammer.

Third, our study only focused on one type of FS (CERAFIT R-MIS) and ancillary and further studies should be performed to evaluate the method for implants manufactured by others.

# 5. Conclusion

This study is the first *ex vivo* validation of the use of an instrumented hammer to assess the insertion endpoint of the FS as well as its primary stability. The results are in agreement with a previous *in vitro* study. The time variation analysis of the force applied by the hammer on the ancillary during the impacts allows the assessment of the number of impacts required for obtaining a satisfactory press-fit condition into the host femoral bone. Further work will focus on the adaptation of this method to clinical conditions, in order to develop a medical device that could help the surgeons estimating the FS stability, with minimal changes of the surgical protocol.

# Acknowledgment

This project has received funding from the European Research Council (ERC) under the European Union's Horizon 2020 research and innovation program (grant agreement No 682001, project ERC Consolidator Grant 2015 BoneImplant).

# Bibliography

Baleani, M., Fognani, R., Toni, A., 2001. Initial Stability of a Cementless Acetabular Cup Design: Experimental Investigation on the Effect of Adding Fins to the Rim of the Cup. Artif. Organs 25, 664–669. https://doi.org/10.1046/j.1525-1594.2001.025008664.x

Barr, J.S., White, J.K., Punt, S.E.W., Conrad, E.U., Ching, R.P., 2015. Effect of Simulated Early Weight Bearing on Micromotion and Pullout Strength of Uncemented Distal Femoral Stems. Orthopedics 38, e417–e422. https://doi.org/10.3928/01477447-20150504-60

Bieger, R., Ignatius, A., Reichel, H., Dürselen, L., 2013. Biomechanics of a short stem: In vitro primary stability and stress shielding of a conservative cementless hip stem. J. Orthop. Res. 31, 1180–1186. https://doi.org/10.1002/jor.22349

Chiba, K., Ito, M., Osaki, M., Uetani, M., Shindo, H., 2011. In vivo structural analysis of subchondral trabecular bone in osteoarthritis of the hip using multi-detector row CT. Osteoarthritis Cartilage 19, 180–185. https://doi.org/10.1016/j.joca.2010.11.002

Cristofolini, L., Varini, E., Pelgreffi, I., Cappello, A., Toni, A., 2006. Device to measure intra-operatively the primary stability of cementless hip stems. Med. Eng. Phys. 28, 475–482. https://doi.org/10.1016/j.medengphy.2005.07.015

Denis, K., Pastrav, L.C., Leuridan, S., 2017. Chapter 20 - Vibration Analysis of the Biomechanical Stability of Total Hip Replacements, in: Zdero, R. (Ed.), Experimental Methods in Orthopaedic Biomechanics. Academic Press, pp. 313–328. https://doi.org/10.1016/B978-0-12-803802-4.00020-2

Fottner, A., Woiczinski, M., Kistler, M., Schröder, C., Schmidutz, T.F., Jansson, V., Schmidutz, F., 2017. Influence of undersized cementless hip stems on primary stability and strain distribution. Arch. Orthop. Trauma Surg. 137, 1435–1441. https://doi.org/10.1007/s00402-017-2784-x

Gortchacow, M., Wettstein, M., Pioletti, D.P., Müller-Gerbl, M., Terrier, A., 2012. Simultaneous and multisite measure of micromotion, subsidence and gap to evaluate femoral stem stability. J. Biomech. 45, 1232–1238. https://doi.org/10.1016/j.jbiomech.2012.01.040

Grechenig, S., Gueorguiev, B., Berner, A., Heiss, P., Müller, M., Nerlich, M., Schmitz, P., 2015. A novel locking screw hip stem to achieve immediate stability in total hip arthroplasty: A biomechanical study. Injury, Current Concepts and Innovations in Trauma Care in Germany 46, S83–S87. https://doi.org/10.1016/S0020-1383(15)30023-1

Henyš, P., Čapek, L., 2018. Impact Force, Polar Gap and Modal Parameters Predict Acetabular Cup Fixation: A Study on a Composite Bone. Ann. Biomed. Eng. 46, 590–604. https://doi.org/10.1007/s10439-018-1980-3

Henyš, P., Leuridan, S., Goossens, Q., Mulier, M., Pastrav, L., Desmet, W., Sloten, J.V., Denis, K., Čapek, L., 2018. Modal frequency and shape curvature as a measure of implant fixation: A computer study on the acetabular cup. Med. Eng. Phys. 60, 30–38. https://doi.org/10.1016/j.medengphy.2018.07.003

Leuridan, S., Goossens, Q., Roosen, J., Pastrav, L., Denis, K., Mulier, M., Desmet, W., Vander Sloten, J., 2017. A biomechanical testing system to determine micromotion between hip implant and femur accounting for deformation of the hip implant: Assessment of the influence of rigid body assumptions on micromotions measurements. Clin. Biomech. Bristol Avon 42, 70–78. https://doi.org/10.1016/j.clinbiomech.2017.01.009

Malfroy Camine, V., Rüdiger, H.A., Pioletti, D.P., Terrier, A., 2018. Effect of a collar on subsidence and local micromotion of cementless femoral stems: in vitro comparative study based on micro-computerised tomography. Int. Orthop. 42, 49–57. https://doi.org/10.1007/s00264-017-3524-0

Mathieu, V., Michel, A., Flouzat Lachaniette, C.-H., Poignard, A., Hernigou, P., Allain, J., Haïat, G., 2013. Variation of the impact duration during the in vitro insertion of acetabular cup implants. Med. Eng. Phys. 35, 1558–1563. https://doi.org/10.1016/j.medengphy.2013.04.005

Mathieu, V., Vayron, R., Richard, G., Lambert, G., Naili, S., Meningaud, J.-P., Haiat, G., 2014. Biomechanical determinants of the stability of dental implants: Influence of the bone–implant interface properties. J. Biomech. 47, 3–13. https://doi.org/10.1016/j.jbiomech.2013.09.021

Michel, A., Bosc, R., Meningaud, J.-P., Hernigou, P., Haiat, G., 2016a. Assessing the Acetabular Cup Implant Primary Stability by Impact Analyses: A Cadaveric Study. PLOS ONE 11, e0166778. https://doi.org/10.1371/journal.pone.0166778

Michel, A., Bosc, R., Sailhan, F., Vayron, R., Haiat, G., 2016b. Ex vivo estimation of cementless acetabular cup stability using an impact hammer. Med. Eng. Phys. 38, 80–86. https://doi.org/10.1016/j.medengphy.2015.10.006

Michel, A., Bosc, R., Vayron, R., Haiat, G., 2015. In Vitro Evaluation of the Acetabular Cup Primary Stability by Impact Analysis. J. Biomech. Eng. 137, 031011. https://doi.org/10.1115/1.4029505

Michel, A., Nguyen, V.-H., Bosc, R., Vayron, R., Hernigou, P., Naili, S., Haiat, G., 2017. Finite element model of the impaction of a press-fitted acetabular cup. Med. Biol. Eng. Comput. 55, 781–791. https://doi.org/10.1007/s11517-016-1545-2

Morohashi, I., Iwase, H., Kanda, A., Sato, T., Homma, Y., Mogami, A., Obayashi, O., Kaneko, K., 2017. Acoustic pattern evaluation during cementless hip arthroplasty surgery may be a new method for predicting complications. SICOT-J 3, 13. https://doi.org/10.1051/sicotj/2016049

Müller, B, Bernhardt, R., Scharnweber, D., Müller, B., Thurner, P., Schliephake, H., Wyss, P., Beckmann, F., Goebbels, J., Worch, H., 2004. Comparison of Microfocus- and Synchrotron X-ray Tomography for the Analysis of Osteointegration around Ti6Al4V Implants. Eur. Cell. Mater. 7, 42–51. https://doi.org/10.22203/eCM.v007a05

Neldam, C.A., Pinholt, E.M., 2014. Synchrotron µCT imaging of bone, titanium implants and bone substitutes – A systematic review of the literature. J. Cranio-Maxillofac. Surg. 42, 801–805. https://doi.org/10.1016/j.jcms.2013.11.015

Nguyen, V.-H., Rosi, G., Naili, S., Michel, A., Raffa, M.-L., Bosc, R., Meningaud, J.-P., Chappard, C., Takano, N., Haiat, G., 2017. Influence of anisotropic bone properties on the biomechanical behavior of the acetabular cup implant: a multiscale finite element study. Comput. Methods Biomech. Biomed. Engin. 20, 1312–1325. https://doi.org/10.1080/10255842.2017.1357703

Oberst, S., Baetz, J., Campbell, G., Lampe, F., Lai, J.C.S., Hoffmann, N., Morlock, M., 2018. Vibro-acoustic and nonlinear analysis of cadavric femoral bone impaction in cavity preparations. MATEC Web Conf. 148, 14007. https://doi.org/10.1051/matecconf/201814814007

Park, Y.-S., Yi, K.-Y., Lee, I.-S., Jung, Y.-C., 2005. Correlation between microtomography and histomorphometry for assessment of implant osseointegration. Clin. Oral Implants Res. 16, 156–160. https://doi.org/10.1111/j.1600-0501.2004.01083.x

Pastrav, L.C., Jaecques, S.V.N., Jonkers, I., Van der Perre, G., Mulier, M., 2009. In vivo evaluation of a vibration analysis technique for the per-operative monitoring of the fixation of hip prostheses. J. Orthop. Surg. 4, 10. https://doi.org/10.1186/1749-799X-4-10

Pastrav, L.C., Jaecques, S.V.N., Mulier, M., Van Der Perre, G., 2008. Detection of the insertion end point of cementless hip prostheses using the comparison between successive frequency response functions. J. Appl. Biomater. Biomech. JABB 6, 23–29.

Sakai, R., Kikuchi, A., Morita, T., Takahira, N., Uchiyama, K., Yamamoto, T., Moriya, M., Uchida, K., Fukushima, K., Tanaka, K., Takaso, M., Itoman, M., Mabuchi, K., 2011. Hammering Sound Frequency Analysis and Prevention of Intraoperative Periprosthetic Fractures during Total Hip Arthroplasty. HIP Int. 21, 718–723. https://doi.org/10.5301/HIP.2011.8823

Schmidutz, F., Woiczinski, M., Kistler, M., Schröder, C., Jansson, V., Fottner, A., 2017. Influence of different sizes of composite femora on the biomechanical behavior of cementless hip prosthesis. Clin. Biomech. 41, 60–65. https://doi.org/10.1016/j.clinbiomech.2016.12.003

Tijou, A., Rosi, G., Hernigou, P., Flouzat-Lachaniette, C.-H., Haïat, G., 2017. Ex Vivo Evaluation of Cementless Acetabular Cup Stability Using Impact Analyses with a Hammer Instrumented with Strain Sensors. Sensors 18, 62. https://doi.org/10.3390/s18010062

Tijou, A., Rosi, G., Vayron, R., Lomami, H.A., Hernigou, P., Flouzat-Lachaniette, C.-H., Haïat, G., 2018. Monitoring cementless femoral stem insertion by impact analyses: An in vitro study. J. Mech. Behav. Biomed. Mater. 88, 102–108. https://doi.org/10.1016/j.jmbbm.2018.08.009

Varini, E., Bialoblocka-Juszczyk, E., Lannocca, M., Cappello, A., Cristofolini, L., 2010. Assessment of implant stability of cementless hip prostheses through the frequency response function of the stem–bone system. Sens. Actuators Phys. 163, 526–532. https://doi.org/10.1016/j.sna.2010.08.029

**Figures captions**

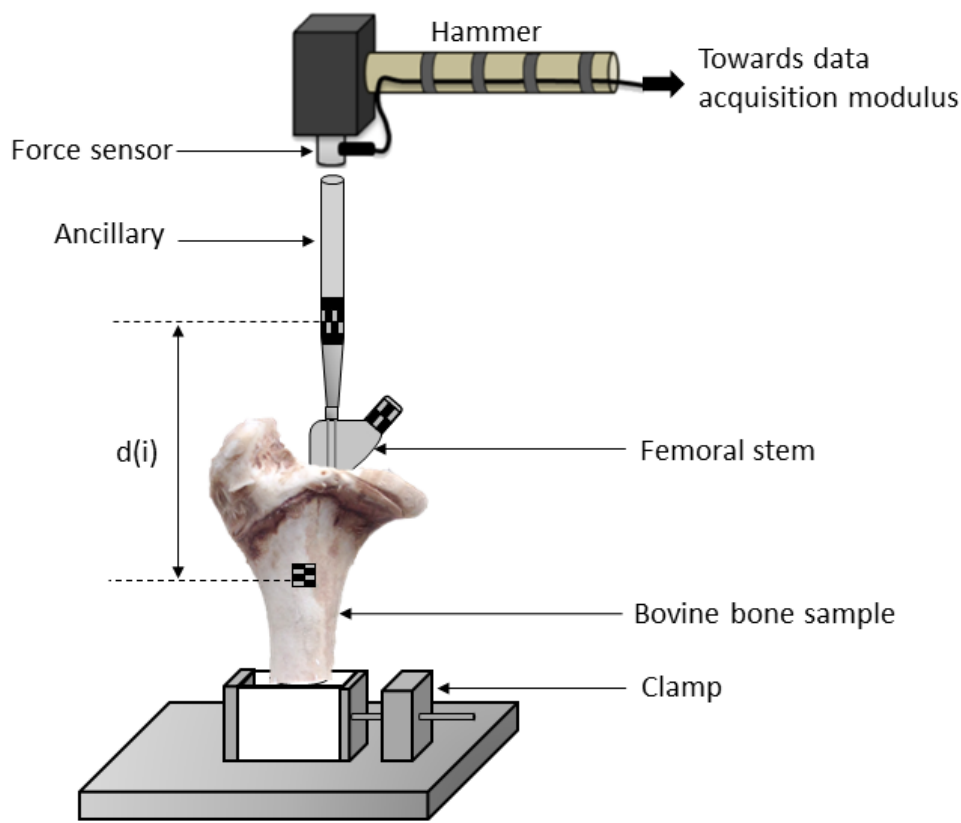

Figure 1 : Schematic representation of the experimental configuration. d(i) corresponds to the distance between a point of the ancillary and of bone tissue after the impact #*i*.

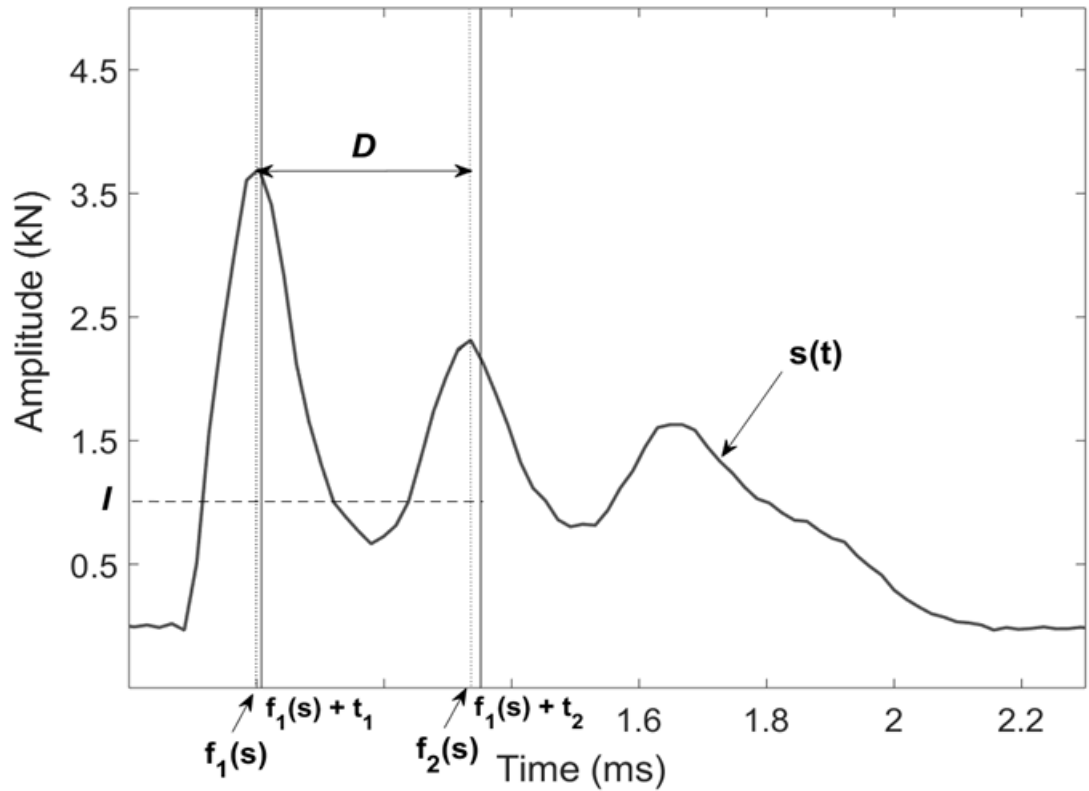

Figure 2 : Illustration of a signal s(t) corresponding to the variation of the force as a function of time obtained for a given impact obtained with an implant size of 7 and a reamer size of 7. The quantities $I$, $f_1(s)+t_1$, $f_1(s)+t_2$, $f_1(s)$, $f_2(s)$, $s(t)$ and $D$ are indicated. The value of $I$ corresponds to average value of $s(t)$ between $f_1(s)+t_1$ and $f_1(s)+t_2$.

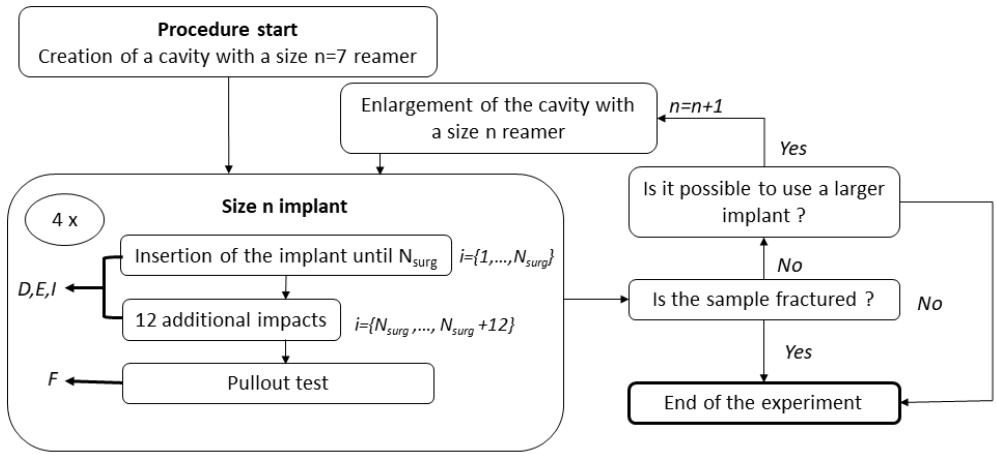

Figure 3 : Schematic representation of the protocol followed for each bovine sample of the first series of experiments corresponding to relatively stable femoral stem

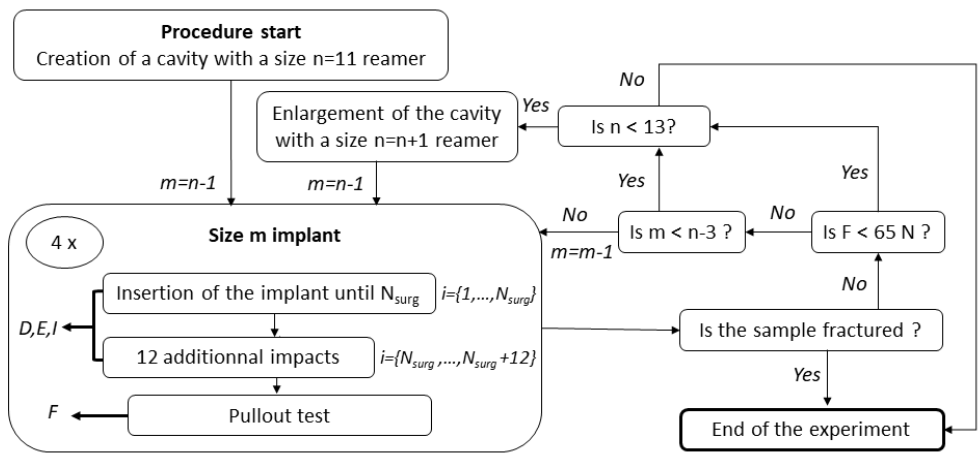

Figure 4 : Schematic representation of the protocol followed for the second series of experiments corresponding to relatively unstable femoral stem

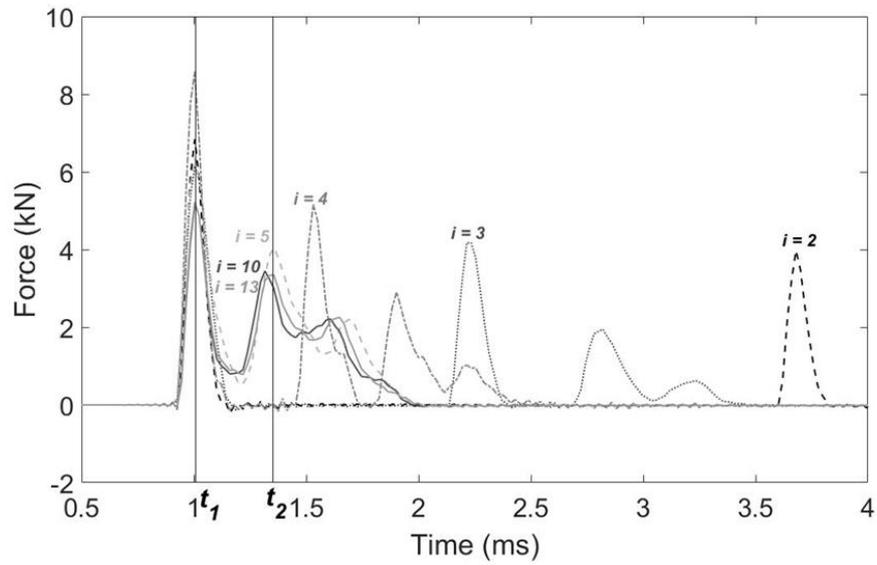

Figure 5 : Time variation of the force applied by the instrumented hammer on the implant- ancillary system for six impacts of sample#4, implant size 9 and test #3. The number $i$ of the impact is indicated on the second peak of each signal. The two vertical lines indicate the time window used for the determination of the indicator $I$.

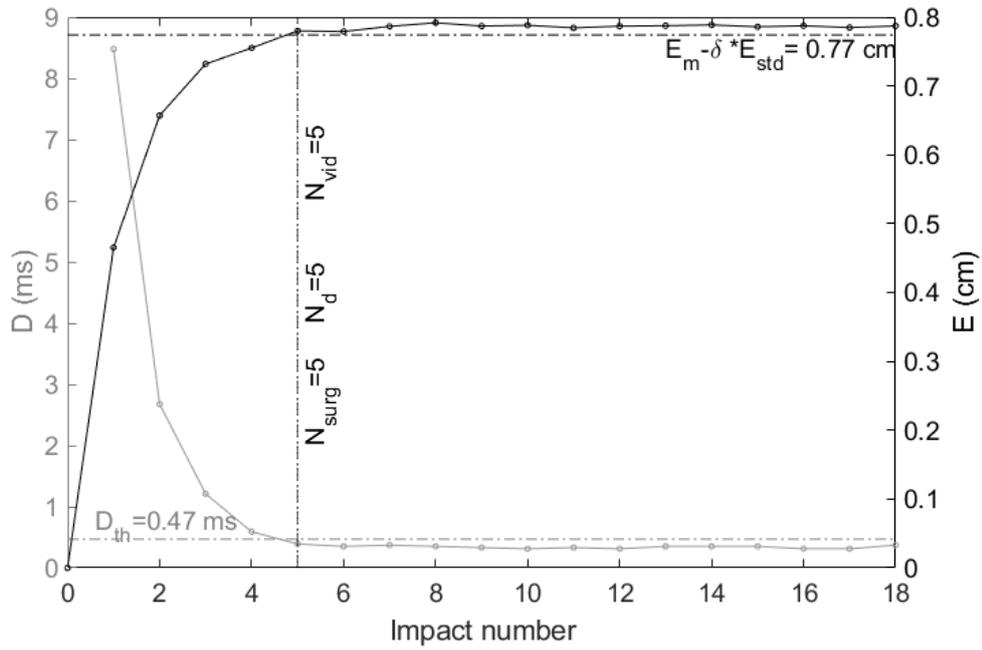

Figure 6 : Variation of D (gray line) and E (black line) indicators as a function of the impact number for the same configuration as the one corresponding to Figure 5. The horizontal black dashed line represents the penetration equal to $E_m - \delta \times E_{std}$ and the horizontal gray dashed line represents the threshold $D_{th} = 0.47\ ms$. The vertical line represent the values obtained for $N_d$, $N_{vid}$ and $N_{surg}$ which are all equal to 5.

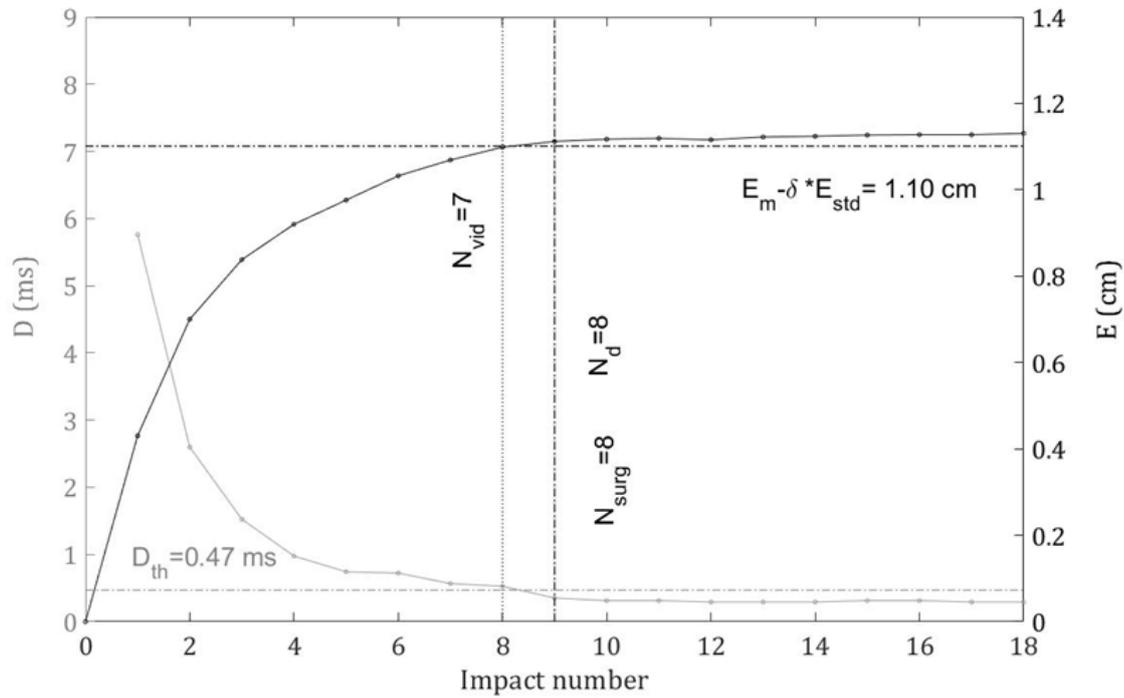

Figure 7: Variation of D (gray line) and E (black line) indicators as a function of the impact number $i$ for the implant size equal to 7 and reamer size equal to 7. The horizontal black dashed line represents the penetration equal to $E_m - \delta \times E_{std}$ and the horizontal gray dashed line represents the threshold $D_{th} = 0.47\ ms$. The vertical black dashed-dotted line represents the value obtained for $N_{surg}$ and $N_d$ which is equal to 8 and the vertical gray dashed line represent the value obtained for $N_{vid}$, which is equal to 7.

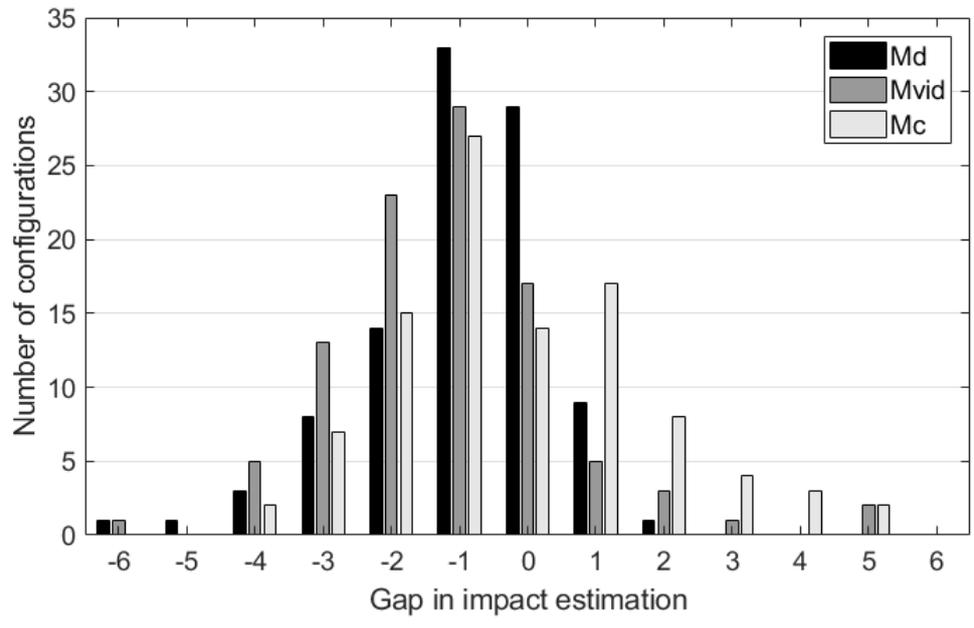

Figure 8 : Distribution of the values obtained for $M_d$, $M_{vid}$ and $M_c$

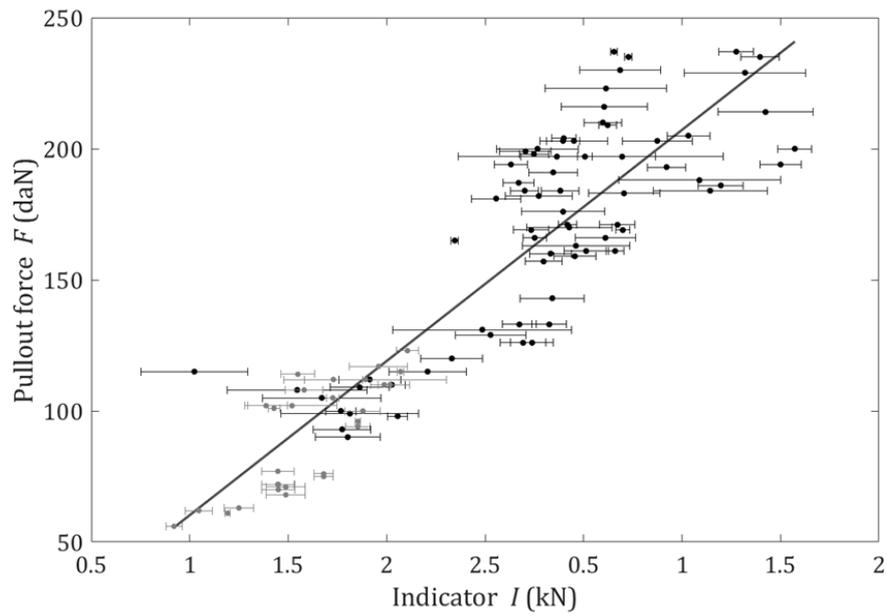

Figure 9 : Relationship between the indicator $I_m$ obtained using the instrumented hammer and the pullout force $F$ ($R^2=0.81$ and $p=1.49 \times 10^{-36}$). The horizontal error bars correspond to the stand deviation values $I_{std;.}$ The black and grey points correspond to the first and second series of experiments, respectively.

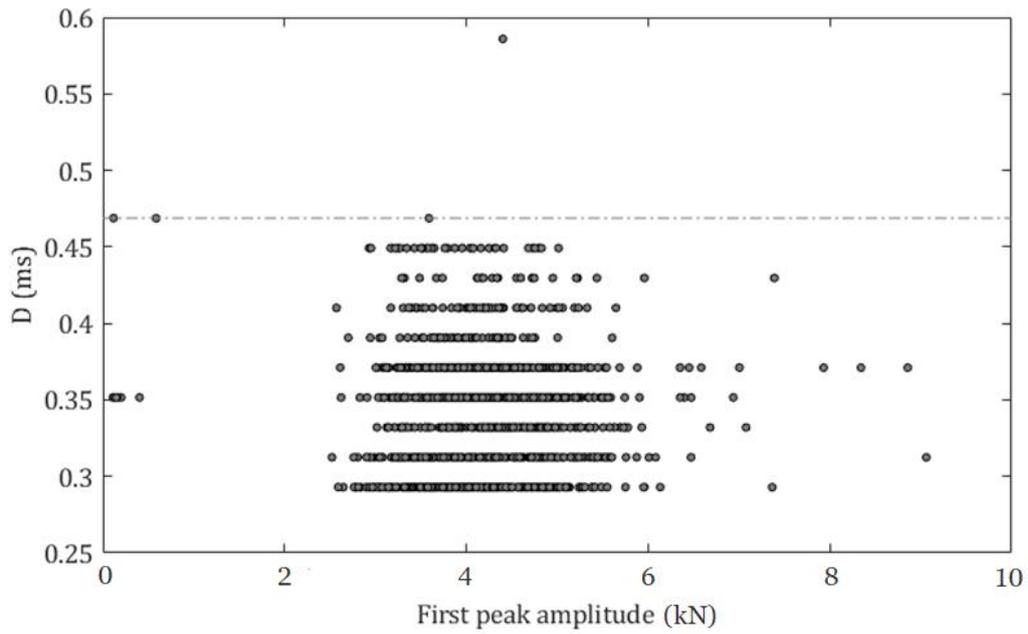

Figure 10 : Variation of the indicator D for all impacts performed after $N_{surg}+1$ for all configurations as a function of the amplitude of the first peak. The horizontal dotted grey line represents the threshold $D_{th}= 0.47$ ms chosen to determine when the femoral stem is fully inserted ($N_d$)